\documentclass[twocolumn,showpacs,preprintnumbers,amsmath,amssymb]{revtex4}
\DeclareMathAlphabet{\mathpzc}{OT1}{pzc}{m}{it}
\usepackage{graphicx}

\begin{document}
%\setcounter{page}{1}
%\preprint{Prepared for Physical Review A}

\title{Dependence of fluorescence-level statistics on bin time size in a few-atom magneto-optical trap}

\author{Sungsam Kang, Seokchan Yoon, Youngwoon Choi, Jai-Hyung Lee, and Kyungwon An}

\email{kwan@phya.snu.ac.kr}
\affiliation{School of Physics, Seoul National University, Seoul 151-742 Korea}

\date{Received \today}

\begin{abstract}
We have analyzed the statistical distribution of the fluorescence signal levels in a magneto-optical trap containing a few atoms and observed that it strongly depends on the relative size of the bin time with respect to the trap decay time. We derived analytic expressions for the signal distributions in two limiting cases, long and short bin time limits, and found good agreement with numerical simulations performed regardless of the size of the bin time. We found an optimal size of the bin time for minimizing the probability of indeterminate atom numbers while providing accurate information on the instantaneous number of atoms in the trap. These theoretical results are compared with actual experimental data. We observed super-Poisson counting statistics for the fluorescence from trapped atoms, which might be attributed to uncorrelated motion of trapped atoms in the inhomogeneous magnetic field in the trap.
\end{abstract}
 \maketitle

\section{INTRODUCTION}
One of the long-sought experimental capabilities in modern atomic
physics and quantum optics is the ability to load a single atom in
a microscopic volume for an extended time and to manipulate and
probe its internal and external states at will. In recent years,
several groups have developed technics for trapping and
controlling a single or a few neutral atoms based on tightly
localized magneto-optical traps or dipole traps
\cite{Ruschewitz96,Hu94,Haubrich96,Youn06}. Single or a-few-atom
traps have been applied to wide range of fields such as cavity
quantum electrodynamics studies \cite{Ye99}, experiments on
single-photon generation on demand \cite{Barquie05}, and even
archeological dating of ancient aquifers \cite{Sturchio04}.

The most distinctive signature of single atom trapping is the quantized
fluorescence signal. When the number of trapped atoms is
decreased to single-atom level, the fluorescence signal from atoms
exhibits stepwise underlying variation in time and the size of the
fluorescence signal with respect to a background level is
interpreted as being proportional to the instantaneous number of
atoms in the trap. Such stepwise fluorescence signals have been
regarded as the most definitive evidence for single atom trapping.
With this understanding, one can obtain the atom-number
distribution in the trap from the histogram of the fluorescence
signal levels and can also identify individual loading and loss
events of atoms in the trap \cite{Youn06,Ueberholz00}.

In actual single-atom trap experiments, since the fluorescence
signal from a single atom is extremely weak, one needs to choose a
bin time for photon counting as long as possible in order to
achieve enough signal to noise ratio. If the bin time is too long,
however, the atom number can change several times during the bin
time and thus the observed fluorescence no longer provides
accurate information on the instantaneous atom number.

From the experimental point of view, therefore, several questions
naturally arise regarding the conditions under which the
fluorescence measurement should be performed: what will be the
optimal size for the bin time, what determines the shape of the
signal distribution and thus what information one can get from the
observed signal distribution. The purpose of this paper
is to answer these questions.

This paper is organized in the following way. In Sec.\ II, we
define the problem and derive analytic expressions for the signal
distributions in two limiting cases, long and short bin time
limits, along with signal-to-noise considerations. These results
are compared with numerical simulations in Sec.\ III, where an
iterative method and Monte Carlo simulations are employed to
calculate steady-state atom number distribution functions and the
signal distributions regardless of the size of the bin time. In
Sec.\ IV, an optimal size of the bin time is identified for
minimizing the probability of indeterminate atom numbers while
providing accurate information on the instantaneous number of
atoms in the trap. The analytic expressions and numerical results
are then compared with experimental results in Sec.\ V. It is
demonstrated that the experimental signal distribution is well fit
by our theoretical model and from observed signal distributions
one can extract information not only on the number of atoms but
also on the state of atoms in the trap. In Sec.\ VI, we summarize
our findings and draw conclusions.

\section{Theoretical Consideration}
In a few-atom trap, the fluorescence signal from atoms, induced by a probe laser or by a weak trap laser itself in the case of a magneto-optic trap (MOT), is proportional to the number of atoms in the trap. The fluorescence signal is measured with a photodetector, usually in photon counting mode with photon counting electronics. Suppose the signal counts are successively taken in time for a preset bin time of $\Delta t$. The signal counts $S_i$ measured in $i^{\rm th}$ time bin, specified as $t_i <t< t_{i+1}$ with $t_i\equiv i\Delta t$ ($i=0, 1, 2, \ldots$), can be written as
\begin{equation}
S_{i}=\int^{t_i+\Delta t}_{t_i} \left(\sum _{j=1}^{N(t)}a_{j}(t)+b(t) \right)dt\;,
\label{eq:signal}
\end{equation}
where $N(t)$ is the instantaneous number of atoms in the trap ($N(t)=0, 1, 2, \ldots$), $a_{j} (t)$ is the counting rate of fluorescence from $j^{\rm th }$ atom, and $b(t)$ is the counting rate of background signal such as detector dark counts and scattered laser light or stray room light. The bin time $\Delta t$ is assumed to be much larger than spontaneous emission lifetime of atoms, typically tens of nanoseconds.
The signal $S_{i}$ is truncated to the nearest integer by the counting electronics.

The instantaneous number of atoms $N(t)$ rapidly fluctuates
due to various stochastic processes. Temporal change of its probability distribution function $P_N(t)$ is governed by the following master equation:
\begin{eqnarray}
\frac{d P_{N}}{dt}&=&R P_{N-1}-\left[R+\Gamma_{1}N+\Gamma_{2}{N\choose 2}\right]P_{N}\nonumber\\
&&+\Gamma_{1}(N+1)P_{N+1}+\Gamma_{2}{N+2\choose2}P_{N+2}\;,
\label{eq:mastereq}
\end{eqnarray}
where $N=0,1,2,\ldots$ with $P_{-1}=0$, $R$ is loading rate of atoms into the trap, $\Gamma_{1}$ is the one-atom loss rate due to collisions with background gas, $\Gamma_{2}$ is the two-atom loss rate due to light-assisted intra-trap collisions \cite{Ueberholz00}. The master equation Eq.(\ref{eq:mastereq}) cannot be solved analytically. However, for a microscopic trap with only a few atoms in a volume of a few micron in diameter, the two-atom loss terms proportional to $\Gamma_2$ are negligibly small, and thus approximate expressions for $P_N$ and the number correlation function $\left<N(t)N(t+\tau)\right>_t$ can be obtained  \cite{Choi06} with $\left< ~ \right>_t$ denoting a time average.
For now, we just neglect the two-atom loss terms. The analysis including these terms will be discussed later.

Without the $\Gamma_{2}$ terms, the master equation Eq.(\ref{eq:mastereq}) becomes the same as the simple birth-death model of population \cite{9}, yielding a Poisson distribution in steady state with a mean value $\bar{N}=R/\Gamma_{1}$ and a variance $\sigma^{2}=\bar{N}$.  For a few-atom trap with $N\sim 1$, the correlation decay time $\tau$ of the number correlation function is given by $\tau= 1/\Gamma_1$ and is the measure of the average time during which $N(t)$ remains constant \cite{Haubrich96}.
When $\Gamma_2$ terms are not negligible,
the correlation decay time is a complicate function of $R$, $\Gamma_1$ and $\Gamma_2$ and always smaller than $1/\Gamma_1$ due to the additional two-atom loss process \cite{Choi06}. We denote the correlation decay time in this case as $\tau_{\rm eff}$ in order to distinguish it from the above definition of $\tau$ for the $\Gamma_2=0$ case.
The correlation decay time is also called the trap decay time for macroscopic traps.

\begin{figure}
\includegraphics[width=3.4in]{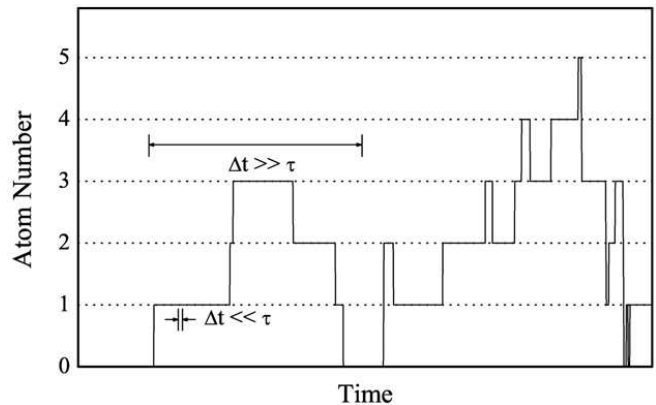}
\caption{Number of atoms in the trap fluctuates in time due to stochastic processes. Two limiting cases of bin time, $\Delta t \gg \tau$ and $\Delta t \ll \tau$, for the integral in Eq.\ (\ref{eq:signal}) are shown. }
\label{fig:example}
\end{figure}

The signal distribution shows much different behaviors depending on the size of $\Delta t$ with respect to $\tau$. For analytic analysis we consider two limiting cases, long bin time ($\Delta t \gg \tau$) and short bin time ($\Delta t \ll \tau$), as illustrated in  Fig. \ref{fig:example}.
In the long bin time limit, there exist many loading and loss events in a single bin time whereas the atom number hardly changes during the bin time in the short bin time limit, and thus these two limiting cases lead to quite different signal distributions.

\subsection{$\Delta t\gg\tau$ limit}

Assume that $N\sim1\ll a_{j}\Delta t, b\Delta t$ and that $a_{j}(t)$ and $b(t)$ fluctuate much faster than $\tau$. Under this assumption, for a given $N$ the variations in $a_{j}\Delta t$ and $b\Delta t$ go like square root of those, respectively, and thus much smaller than $a_j \Delta t$ and $b\Delta t$ themselves. On the other hand, the variation due to $N$ change is as large as $a_j \Delta t$. Therefore, in evaluating the integral in Eq.\ (\ref{eq:signal}), we can neglect the fluctuations in $a_{j}$ and $b$ and replace them with their mean values $\bar{a}$ and $\bar b$, respectively.
\begin{equation}
S_{i}\simeq\int_{t_i}^{t_i+\Delta t}[N(t)\bar{a}+\bar{b}]dt=\left(N_{i} \bar{a}+\bar{b}\right)\Delta t \;,
\label{eq3}
\end{equation}
where
\begin{equation}
N_{i}=\frac{1}{\Delta t}\int_{t_i}^{t_i+\Delta t}N(t)dt
\end{equation}
is the time-averaged atom number in the $i^{\rm th}$ time bin. Since $\Delta t \gg\tau$, $N_{i}$, which is no longer an integer, fluctuates around the mean value $\bar{N}$ with a new variance $\tilde{\sigma}^{2}$, which is not the same as the variance $\sigma^2$ of $P_N$ distribution above. In fact, $\tilde{\sigma}^{2}\rightarrow 0$ as $\Delta t\rightarrow \infty$.
The probability distribution $P(N_{i})$ can be obtained by the central limit theorem as a Gaussian distribution,
\begin{equation}
P(N_{i})=\frac{1}{\sqrt{2\pi}\tilde{\sigma}}\exp\left[-\frac{1}{2\tilde{\sigma}^{2}}(N_{i}-\bar{N})^{2}\right]\;,
\label{eq:Pnlarget}
\end{equation}
where the variance $\tilde{\sigma}^2$ is proportional to the original variance $\sigma^2$, which equals $\bar{N}$ for a Poisson distribution, and inversely proportional to the sample size, which is in the order of $\Delta t/\tau$.
The exact calculation for a Poisson distribution ($\Gamma_2=0$ case) is given below:
\begin{eqnarray}
\tilde{\sigma}^2&=&\langle N_{i}^{2}\rangle - \langle N_{i}\rangle^{2}\nonumber\\
&=&\frac{1}{\Delta t^{2}}\left< \int_{t_i}^{t_i+\Delta t}dt \int_{t_i}^{t_i+\Delta t}dt'N(t)N(t')\right>-\bar{N}^{2}\nonumber\\
&=&\frac{1}{\Delta t^{2}}\int_{0}^{\Delta t}dt\int_{0}^{\Delta t}dt'\left< N(t)N(t')\right>_{t}-\bar{N}^{2}\;,\nonumber\\
\end{eqnarray}
where $\langle~\rangle$ denotes an ensemble average and by the ergodic theorem the ensemble average is replaced with the time average. Since the correlation function for a Poisson distribution is given by \cite{9}
\begin{equation}
\langle N(t)N(t')\rangle_{t}=\bar{N}^{2}+\bar{N}e^{-|t'-t|/\tau}\;,
\label{eq:correlation}
\end{equation}
the variance becomes
\begin{equation}
\tilde{\sigma}^2=\frac{2\bar{N}\tau^2}{\Delta t^{2}}
\left(e^{-\Delta t/\tau}+\Delta t/\tau-1\right)\;.
\label{eq:variance}
\end{equation}
In the limit of $\Delta t\gg\tau$, we then obtain
\begin{equation}
\tilde{\sigma}^2 \simeq 2\bar{N}\tau/\Delta t
\label{eq9}
\end{equation}
as expected.

\begin{figure*}[t]
\includegraphics[width=5.0in]{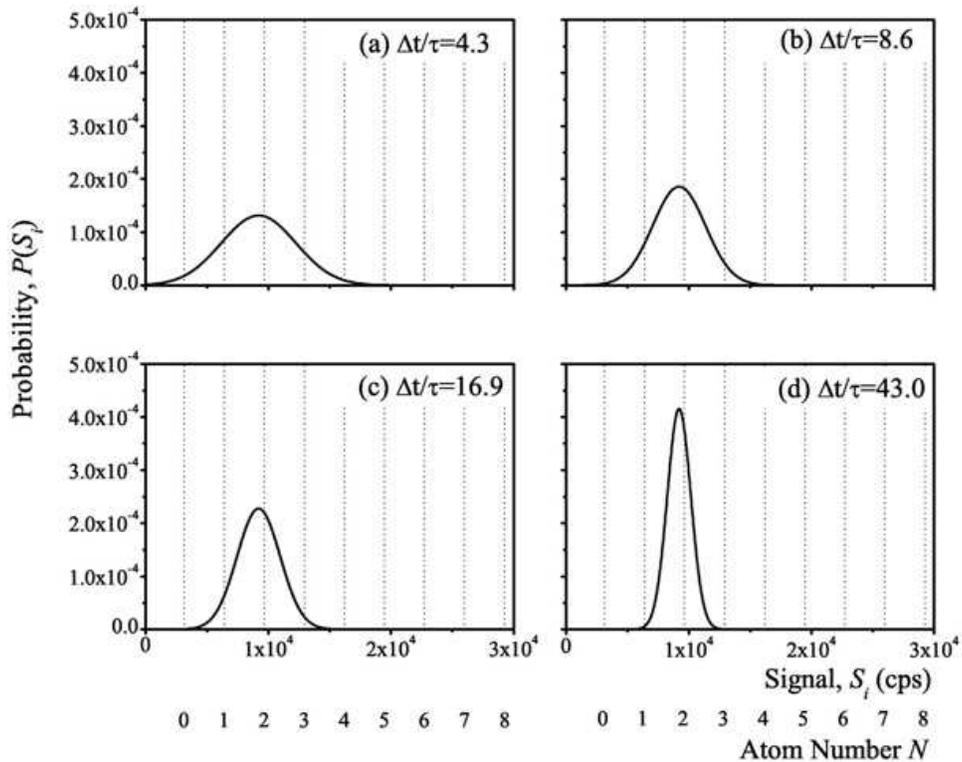}
\caption{Signal distribution given by Eq.\ (\ref{eq:Pslarget}) with $\bar{N}$=1.9, $\bar{a}$=3270 s$^{-1}$, $\bar{b}$=3140 s$^{-1}$ and $\tau$=23 s. These parameters are derived from the experimental data in
Ref.\ \cite{Choi06}. In plotting the distributions, signal counts
are divided by corresponding bin times for the horizontal axis so
as to align the distributions with different bin times.  The
dotted lines indicate corresponding atom numbers.}
\label{fig:singlepeak}
\end{figure*}

 The probability distribution for $S_{i}$ is then
obtained from Eqs.\ (\ref{eq3}) and (\ref{eq:Pnlarget}) as
\begin{equation}
P(S_i)=\frac{1}{\sqrt{2\pi}\sigma_S}\exp\left[-\frac{1}{2\sigma_S^2}(S_i-\bar{S})^2\right]\;,
\label{eq:Pslarget}
\end{equation}
where
\begin{equation}
\bar{S}=(\bar{N}\bar{a}+\bar{b})\Delta t
\end{equation}
and the deviation $\sigma_S$ is given by
\begin{equation}
\sigma_S=\tilde{\sigma}\bar{a}\Delta t \rightarrow \sqrt{2\bar{N}\bar{a}^2\tau \Delta t}
\end{equation}
with the arrow indicating approximation under the condition of $\Delta t/\tau \rightarrow \infty$ and $\Gamma_2=0$.
The signal to noise ratio $(S/N)_S$ for $S_i$ is given by
\begin{eqnarray}
(S/N)_S &\equiv& \frac{\bar{S}}{\sigma_{S}}
=\frac{(\bar{N}\bar{a}+\bar{b})}
{\tilde{\sigma}\bar{a}} \nonumber\\
&\rightarrow& [\bar{N}+\bar{b}/\bar{a}]\sqrt{\frac{\Delta t}
{2\bar{N}\tau}}
\propto \sqrt{\Delta t}\;,
\label{eq:snr_long}
\end{eqnarray}
increasing as the square root of the bin time. Fig.\
\ref{fig:singlepeak} shows the behavior of $P(S_{i})$ for
different ratio of $\Delta t/\tau$. The distribution is
single-peaked centered around $\bar{S}$ and the relative width of
the peak with respect to the mean $\bar{S}$ becomes narrower as
$\Delta t$ is made larger, as expected from
Eq.(\ref{eq:snr_long}).

\subsection{$\Delta t\ll\tau$ limit}
In this limit, $N(t)$ remains at a certain integer value $m$ through out the bin time and thus
\begin{eqnarray}
S_{i}&=&\int^{t_i+\Delta t}_{t_i}\left(\sum_{j=1}^{m}a_{j}(t)+b(t)\right) dt\nonumber\\
&=&\sum_{j=1}^m A_{j, i}+B_i\;,
\label{eq:aab}
\end{eqnarray}
where
\begin{eqnarray}
A_{j, i}&=&\int_{t_i}^{t_i+\Delta t}a_{j}(t) dt\nonumber\\
B_i&=& \int_{t_i}^{t_i+\Delta t}b(t)dt
\end{eqnarray}
are the number of fluorescence and background counts in the bin time, respectively, and thus integers.

In general, the statistics of $A_{j,i}$ and $B_i$, with associated
distribution functions $P_a(A_{j,i})$ and $P_b(B_i)$,
respectively, are not necessarily Poissonian. However, in many
cases these quantities follow Poisson statistics. For example,
although the photon statistics of resonance fluorescence of a
small number of atoms is sub-Poissonian, when measured with an
imperfect photodetector, the counting statistics become
Poissonian. In addition, statistics of scattered light of laser
beam is Poissonian. Of course, there are cases where these
statistics become super-Poissonian, particularly when laser power
fluctuations and other technical noises enter. For now, we just
assume both $A_{j,i}$ and $B_i$ follow Poisson statistics with
mean values $\bar{A}\equiv\bar a \Delta t$ and
$\bar{B}\equiv\bar{b}\Delta t$, respectively.

Under this assumption, the conditional probability for $S_{i}$ with a constraint $N=m$ is given by

\begin{equation}
P(S_i|m)={\sum}'_{\left\{A_{j, i},B_i\right\}}\left[\prod_{j=1}^m P_a(A_{j,i})\right] P_b(B_i)
\end{equation}
where ${\sum}'_{\left\{A_{j, i},B_i\right\}}$ represents summations to be performed for all possible combinations of $A_{j, i}$ and $B_i$ under the constraint of Eq.\ (\ref{eq:aab}). If we assume Poisson distributions for $P_a(A_{j,i})$ and $P_b(B_i)$,
\begin{eqnarray}
P(S_i|m)
&=&{\sum}'_{\left\{A_{j, i},B_i\right\}}
\left[ \prod_{j=1}^m \frac{\bar{A}^{A_{j,i}} e^{-\bar{A}}}{A_{j,i}!} \right]
\frac{\bar{B}^{B_i} e^{-\bar{B}}}{B_i!}\nonumber\\
&=&\frac{\left(e^{-\bar{A}}\right)^me^{-\bar{B}}}{S_i!}{\sum}'_{\left\{A_{j, i},B_i\right\}}S_i!
\left[ \prod_{j=1}^m \frac{\bar{A}^{A_{j,i}}}{A_{j,i}!} \right]
\frac{\bar{B}^{B_i} }{B_i!}\nonumber\\
&=& \frac{e^{-\bar{S}_m}}{S_i!}\left[\left(\sum_{j=1}^m \bar{A}\right)+\bar{B}\right]^{S_i}\nonumber\\
&=& \frac{\bar{S}_m^{S_i}}{S_i!}e^{-\bar{S}_m}
\label{eq:sm}
\end{eqnarray}
where
\begin{equation}
\bar {S}_{m}=m\bar{A}+\bar{B}\;.
\label{eq:sbar}
\end{equation}
The resulting distribution is just a Poisson distribution with both mean value and variance equal to $\bar{S}_{m}$. For non-Poisson distributions for  $P_a(A_{j,i})$ and $P_b(B_i)$, the resulting $P(S_i|m)$ is not Poissonian. However, it is still a well-localized Gaussian-like distribution with a mean value $\bar{S}_m$, but its variance is no longer equal to $\bar{S}_m$.

\begin{figure}
\includegraphics[width=3.4in]{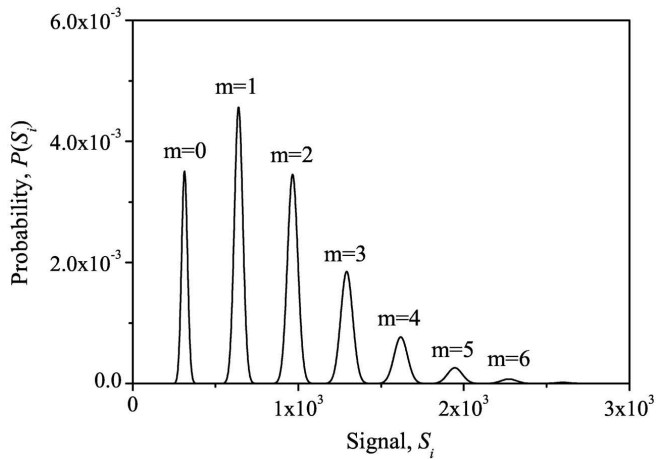}
\caption{Signal distribution given by Eq.\ (\ref{eq:smallt}) with $\bar{N}$=1.9, $\bar{a}$=3270 s$^{-1}$, $\bar{b}$=3140 s$^{-1}$, $\tau$=23 s, $\Delta t$=0.100 s, and thus $\Delta t / \tau$=0.0043. These
parameters are derived from the experimental data in Ref.\
\cite{Choi06}.
Each peak in the distribution is marked with a corresponding atom number $m$.} 
\label{fig:multipeak}
\end{figure}

The probability distribution for $S_{i}$ for all possible $N$ values is then given by
\begin{equation}
P(S_{i})=\sum_{m=0}^{\infty}P(S_{i}|N=m)\cdot P_{N=m}
\label{eq:smallt}
\end{equation}
where each $P(S|m)$, peaked around its mean $\bar{S}_m$ with a variance $\sigma_m^2$, is modulated by $P_{m}$ as shown in Fig.\ \ref{fig:multipeak}.
The signal to noise ratio $(S/N)_m$ for the $N=m$ signal level becomes,
\begin{equation}
(S/N)_m=\sqrt{\bar{S}_{m}}=\sqrt{(m\bar{a}+\bar{b})\Delta t}\propto \sqrt{\Delta t}\;.
\end{equation}
The half width of the $m$th peak in the signal distribution is also given by the square root of $\bar{S}_m$ and thus the ratio of the $m$th-peak full width
to the spacing of two adjacent peaks is equal to
\begin{equation}
2\sqrt{\bar{S}_{m}}/\bar{A}=2\sqrt{\frac{m+\bar{b}/\bar{a}}{\bar{a} \Delta t}}\;.
\label{eq:snr_short}
\end{equation}
Unless this ratio is very small, the adjacent peaks substantially overlap and thus we have a significant probability of indeterminate atom numbers. 
A necessary condition for well separated adjacent peaks is then $\tau\gg \Delta t \ggg 4/\bar{a}$.

For example, consider the set of parameters used in Figs.\ 2 and 3, $\bar{a}$=3270 s$^{-1}$,  $\bar{b}$=3140 s$^{-1}$ and $\tau$=23 s, but with a very short bin time, $\Delta t/\tau$=0.00043. For these parameters the ratio in Eq.\ (21) is 0.34, 0.48, 0.59, 0.68 for $m=0,1,2,3$, respectively, and thereby results in significant overlap between adjacent peaks for $m \ge 1$. This situation is illustrated in Fig.\ 6(a), which is the result of Monte Carlo simulation obtained for this set of parameters. Detailed discussion on Fig.\ 6 will be given in the next section.

\section{Numerical Simulations}
In the preceding sections, we argued that the two-atom loss terms
in the master equation are negligibly small for a few-atom trap
with a few-micron in size and thus the atom number distribution
function is approximately Poissonian. When the number of atoms in
such microscopic trap is increased with its size fixed, the
two-atom loss processes take place more frequently. As a result,
the atom-number distribution deviates significantly from a
Poissonian distribution and thus the Poisson approximation in the
preceding sections are no longer applicable. In this section, we
include the two-atom loss term and calculate distribution
functions numerically.

Although the master equation, Eq.\ (\ref{eq:mastereq}), cannot be
solved analytically, a steady-state solution can be found
numerically. In steady state, we have $dP_{N}/dt=0$ and by
rearranging terms we obtain the following recursion relation with
$P_{-1}=0$.
\begin{eqnarray}
RP_{N}&=&\frac{1}{2}(N+1)(2\Gamma_{1}+N \Gamma_{2})P_{N+1}\nonumber\\
& &+\frac{1}{2}(N+2)(N+1)\Gamma_{2}P_{N+2}.\label{eq:mastereq3}
\end{eqnarray}
Using this relation the atom-number distribution $P_N$ can be
easily calculated by iterative method. Alternatively, one can
calculate a fluctuating time sequence of $S_i$ in steady state by
simulating loading and one- and two-atom losses and simulating
fluctuating $a_j(t)$ and $b(t)$ in Monte Carlo simulation. From
the time sequence, one can calculate the histogram of atom number,
{\em i.e.}, the steady-state atom-number distribution.

\begin{figure}[t!]
\includegraphics[width=3.4in]{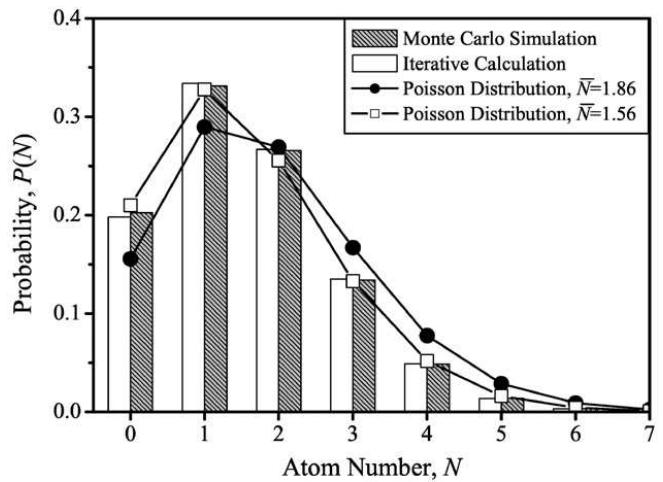}
\caption{Atom number distribution for $R$=0.080 s$^{-1}$,
$\Gamma_{1}$=0.043 s$^{-1}$ and $\Gamma_{2}$=0.0056 s$^{-1}$. For
Monte Carlo simulation, we assumed $\Delta t$=0.20 s. Empty bar
shows the probability calculated by iterative method and the
filled bar that  by Monte Carlo simulation. Filled circle-line
shows a Poisson distribution with the same $R$ and $\Gamma_{1}$
but $\Gamma_2=0$ whereas empty square-line represents a Poisson
distribution with the same mean value as the iterative and Monte
Carlo simulations.}. \label{fig:pn}
\end{figure}

We compare the results of these two numerical methods in Fig.\
\ref{fig:pn}. The values of $R, \Gamma_1$ and $\Gamma_2$ used in
the calculations were derived from the experimental data of Ref.\
\cite{Choi06}.
A Poisson distribution with the same $R$ and
$\Gamma_1$ is also shown in Fig.\ \ref{fig:pn} (by filled circle-line) for comparison.
Once $P(N)$ is known, we can calculate the mean atom number
$\bar{N}$ and variance $\sigma^2$. The results are $\bar{N}=%1.56
1.6$
and $\sigma^{2}=%1.45
1.5$, which should be compared with
$\bar{N}=\sigma^{2}=R/\Gamma_1=%1.86
1.9$ obtained for $\Gamma_{2}=0$.
With inclusion of $\Gamma_2$ term, the mean atom number decreases
because of the additional loss term. Although the distribution is
not Poissonian, the deviation from a Poissonian distribution with
the same $\bar{N}$ value is negligibly small. According to
Ref.\cite{Choi06}, the correlation function which includes the two
atom loss term can be approximated by the functional form for Poisson case as in Eq.\ (\ref{eq:correlation})
with $\tau$ replaced with an effective correlation decay time $\tau_{\rm eff}$.
\begin{equation}
\tau_{\rm eff}=1/\Gamma_{\rm
eff}=4/[\Gamma_{1}+3\sqrt{\Gamma_{1}^{2}+4R\Gamma_{2}}] \nonumber
\end{equation}
For the above parameters, $\tau_{\rm eff}$=%16.6
18 s, compared to $\tau= \Gamma_1^{-1}$=%23.3
23 s.

\begin{figure}[h!]
\includegraphics[width=3.4in]{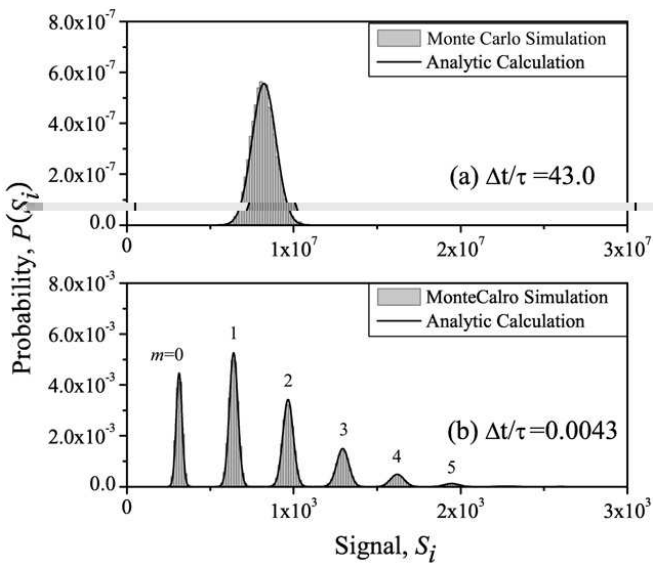}
\caption{Signal distribution for $R$=0.080 s$^{-1}$,
$\Gamma_{1}$=0.043 s$^{-1}$, $\Gamma_{2}$=0.0056 s$^{-1}$,
$\bar{a}$=3270 s$^{-1}$ and $\bar{b}$=3140 s$^{-1}$. (a) $\Delta t
\gg\tau$ limit, and (b) $\Delta t \ll\tau$ limit. Solid curves in
(a) and (b) are obtained by Eq.(\ref{eq:Pslarget}) and
Eq.(\ref{eq:smallt}), respectively, and filled areas are the
result by Monte Carlo simulations in both cases. In (b), each peak is marked with a corresponding atom number.}
\label{fig:compare}
\end{figure}

This observation allows us to use Eqs.(\ref{eq:Pnlarget}) and
(\ref{eq:Pslarget}) for the calculation of $P(N_i)$ and $P(S_i)$,
respectively, with the $\bar{N}$ and $\tau_{\rm eff}$ values
obtained above for nonzero $\Gamma_2$. For $\Delta t\ll\tau$
limit, we can calculate $P(S_i)$ distribution by using Eq.\
(\ref{eq:smallt}) with substitution of the exact $P_{N=m}$
obtained numerically. In Fig.\ \ref{fig:compare} the solid lines
are given by Eqs.\ (\ref{eq:Pslarget}) and (\ref{eq:smallt}) and
the filled area is by the Monte Carlo simulation.

\begin{figure*}
\includegraphics[width=5in]{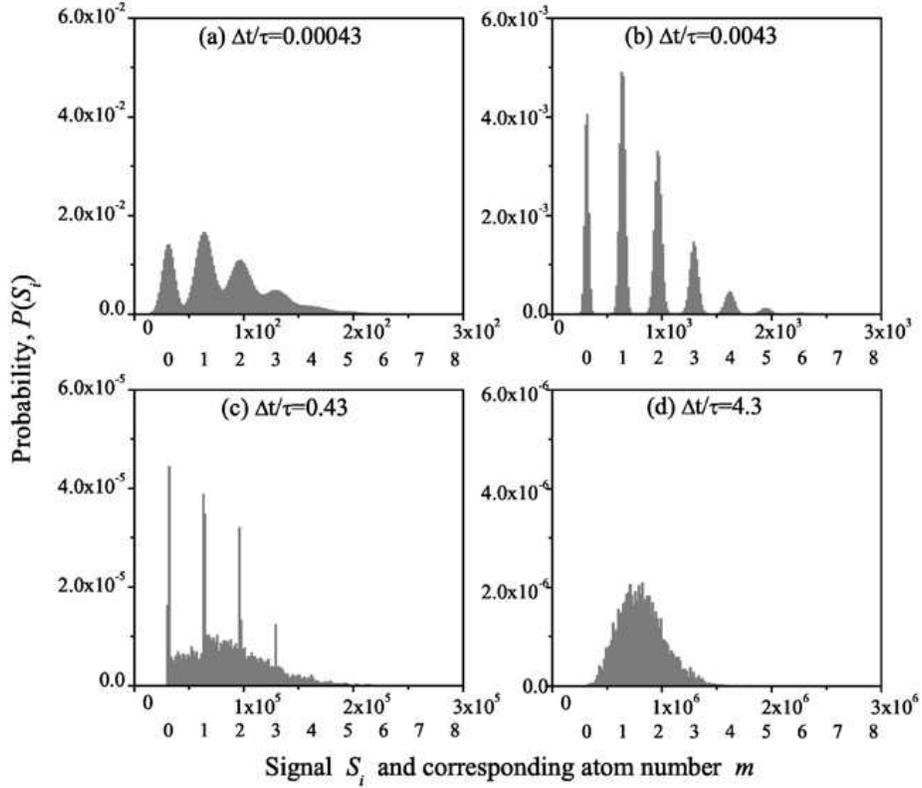}
\caption{Dependence of signal distributions on $\Delta t/\tau$.
(a), (b), (c) and (d) shows the results of Monte Carlo
simulations. } \label{fig:inter}
\end{figure*}

The signal distribution $P(S_i)$ in the intermediate region, other
than two limiting cases considered above, can only be obtained by
Monte Carlo simulations.
%\subsection{This part should be rewritten}
From the time sequence of $S_i$ calculated by means of the Monte
Carlo simulation with the aforementioned parameters, we can
calculate $P(S_i)$ for various $\Delta t/\tau$ by combining $S_i$
values in neighboring time bins. The results are summarized in
Fig.\ \ref{fig:inter}. For $\Delta t/\tau \ll 1$, individual
atom-number peaks are well separated and resolved as shown in
Fig.\ \ref{fig:inter}(b) as long as $\Delta t/\tau \ggg
4/(\bar{a}\tau)$. Otherwise, the peaks for large $m$ overlap with
neighboring peaks significantly as shown in Fig.\
\ref{fig:inter}(a). As the ratio $\Delta t/\tau$ increases, the
broad background appears and grows in height as in Fig.\
\ref{fig:inter}(c) until the background outgrows the atom-number
peaks completely as in Fig.\ \ref{fig:inter}(d).

\section{Optimal Bin Time}
The trend observed in Fig.\ \ref{fig:inter} can be formulated in a quantitative way.
We have observed for $\Delta t \ll \tau$ that individual signal distributions significantly overlap with neighboring peaks (due to poor signal-to-noise ratio) unless $\Delta t$ is much greater than $4/\bar{a}$. The overlap of distribution functions leads to an increase in the probability of having indeterminate atom numbers.
We can quantify this probability ${\mathpzc P}_<$ as a sum of all areas under the distribution function $P(S_i)$ outside the boundaries set by
$\bar{S}_m - \eta\bar{A} <S_i<\bar{S}_m + \eta\bar{A}$ around the $m$th peak with $\eta <0.5$. In the time trace picture of fluorescence signal as in Fig.\ 7, this probability is proportional to the number of data points outside the region specified by dotted lines around a mean signal level. For these data points the atom number cannot be assigned unambiguously.
From Eq.\ (\ref{eq:smallt}) we then obtain
\begin{equation}
{\mathpzc P}_<(\Delta t)=\frac{1}{1-2\eta}\left[1-\sum^{\infty}_{m=0}\int^{\bar{S}_{m}+\eta\bar{A}}_{\bar{S}_{m}-\eta\bar{A}}P(S_{i})\ dS_{i}\right]
\end{equation}
where the factor $1/(1-2\eta)$ is introduced in order to make $\mathpzc{P}_<$ be properly normalized in the limit of $\eta\rightarrow 0$.

\begin{figure}
\includegraphics[width=3.4in]{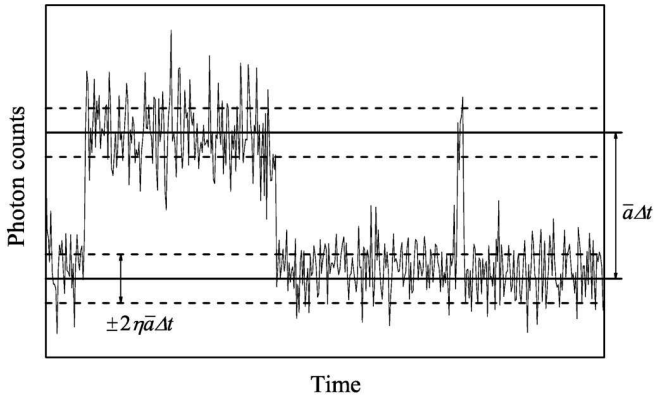}
\caption{Signal counts outside a region centered around the mean
signal level cannot be associated unambiguously with the atom
number corresponding to the mean signal level.} \label{fig:inter0}
\end{figure}

The atom number also becomes indeterminate if it changes during the bin time as in the case of $\Delta t \gg \tau$. From the master equation, Eq.\ (\ref{eq:mastereq}), it can be seen that the total rate of change $\Gamma_{\rm tot}$ of the atom number is given by
\begin{equation}
\Gamma_{\rm tot}(m)=R+\Gamma_{1}m+\Gamma_{2}{m\choose 2}.
\end{equation}
for the atom number $m$ at that instance.
The probability that the atom number would change from $N=m$ during $\Delta t$ is then
\begin{equation}
{\mathpzc P}(\Delta t|N=m)=1-\exp[-\Gamma_{\rm tot}(m)\Delta t].
\end{equation}
By summing over all possible atom numbers according to $P_N$, we obtain the probability ${\mathpzc P}_>(\Delta t)$ that the atom number would change during $\Delta t$ regardless of its initial values.
\begin{equation}
{\mathpzc P}_>(\Delta t)=\sum^{\infty}_{N=0}\left\{ 1-\exp \left[ -\Gamma_{\rm tot}(N)\Delta t \right] \right\} \cdot P_N.
\label{eq:changing_prob2}
\end{equation}
If the atom number changes during the bin time $\Delta t$, the atom number cannot be determined unambiguously from the signal level for this particular bin time. Therefore, ${\mathpzc P}_>(\Delta t)$ can be regarded as the total probability of indeterminate atom numbers for $\Delta t \gtrsim \tau$.

\begin{figure}
\includegraphics[width=3.4in]{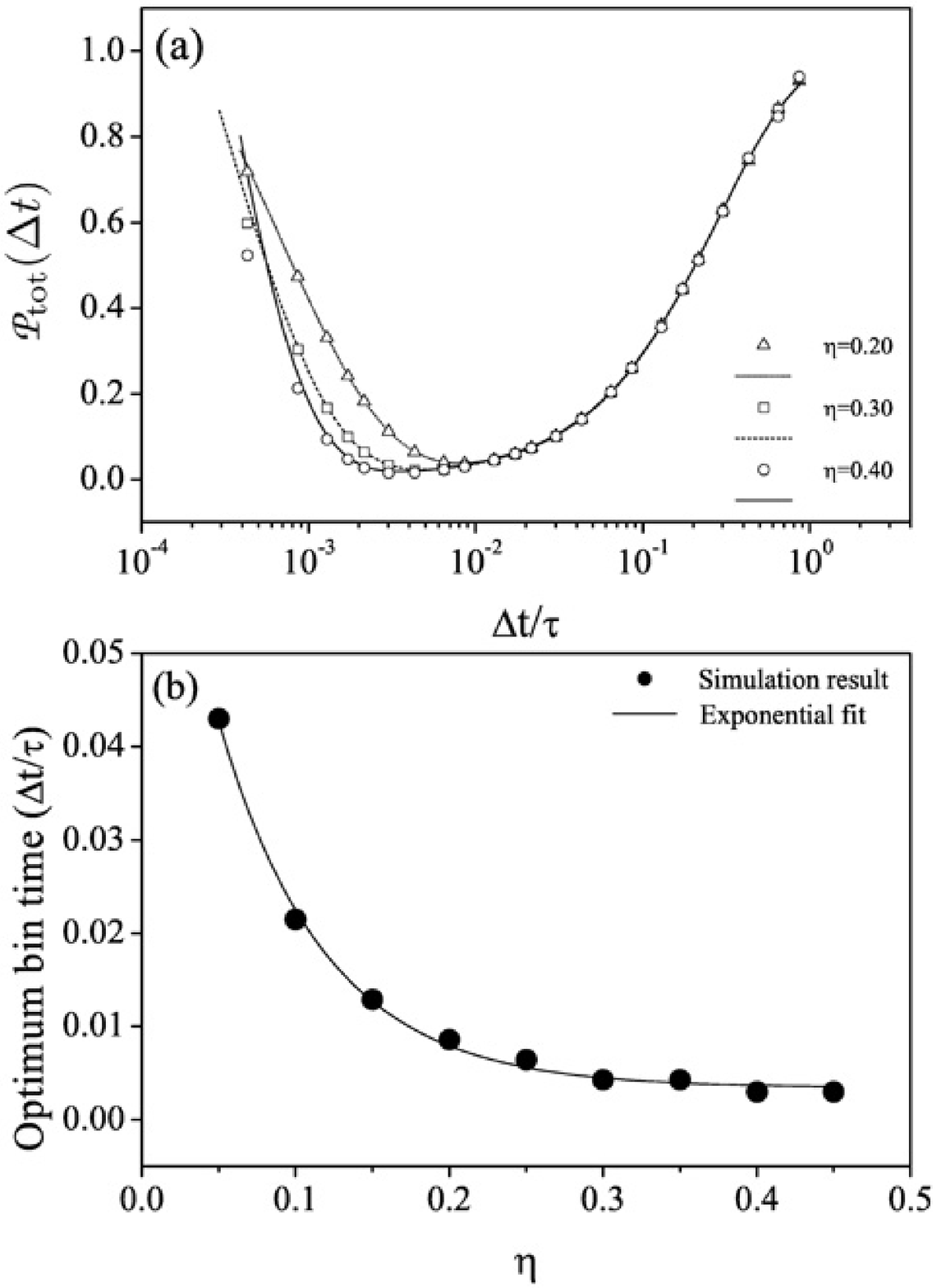}
\caption{(a) Total probability ${\mathpzc P}_{\rm tot}(\Delta t)$ of indeterminate atom numbers, given by Eq.\ (\ref{eq27}), for various values of $\eta$.  Symbols
denote Monte-Carlo simulation results. (b) Optimal bin time as a
function of $\eta$.} \label{fig7}
\end{figure}

In general, the above two processes occur independently and thus can occur simultaneous during $\Delta t$. Therefore, the total probability of indeterminate atom numbers for arbitrary $\Delta t$ is given by
\begin{equation}
{\mathpzc P}_{\rm tot}(\Delta t)={\mathpzc P}_>(\Delta t)+{\mathpzc P}_<(\Delta t)-{\mathpzc P}_>(\Delta t)\cdot {\mathpzc P}_<(\Delta t).
\label{eq27}
\end{equation}
In Fig.\ \ref{fig7}(a), this probability ${\mathpzc P}_{\rm
tot}(\Delta t)$ is plotted as the ratio $\Delta t/\tau$ for
several $\eta$ values. Symbols represent the results of
Monte-Carlo simulations. The bin time $\Delta t$ that minimizes
this probability can be regarded as an optimal bin time for
accurate measurement of the instantaneous atom numbers in a
few-atom trap. The optimal $\Delta t/\tau$ value is plotted as a
function of $\eta$ in Fig.\ \ref{fig7}(b). It can be seen that for
$0.2<\eta<0.4$ the optimal bin time is within the range of $0.003
<\Delta t/\tau <0.008$. Particularly, for $\eta=0.3$, we have an
optimal bin time of $\Delta t/\tau=0.004$ for the same parameter
values as used in Figs.\ 2--6.
Among the plots in Fig.\ 6, plot (b) is the most closest to the case of the optimal bin time.

\section{Comparison with Experimental Data} \label{sec4}
\begin{figure}[t!]
\includegraphics[width=3.4in]{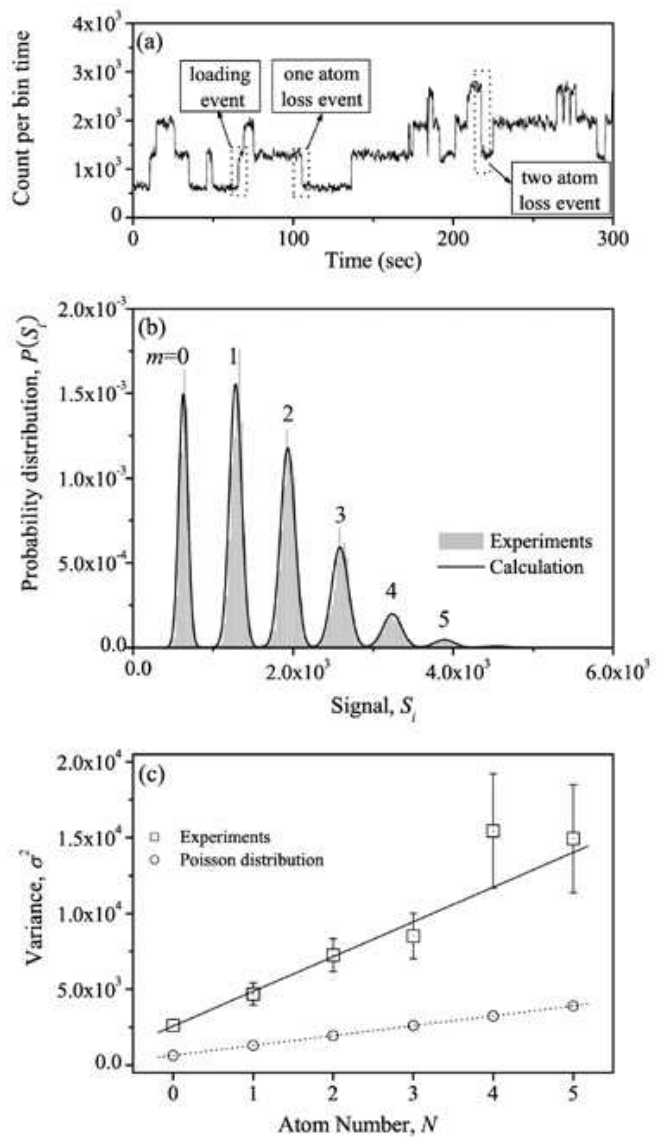}
\caption{(a) A segment of the fluorescence signal time trace
observed in the experiment in Ref.\ \cite{Choi06}. Loading events
and atom- and two-atom loss events can be identified in the trace.
(b) Signal distribution obtained from the experimental data in
(a). The filled area shows the experimental result.
The solid line shows a fit based on Eqs.\ (\ref{eq:smallt}) and
(\ref{eq:variance_m_2}). Each peak is marked with a corresponding atom number. (c) The filled squares represent the
variances of individual atom-number peaks in the observed signal
distribution in (b) whereas the unfilled squares show those of
Poisson statistics. Bin time $\Delta t$ was 0.20 s and the correlation
decay time $\tau$ was about 23 s in both experiment and analysis.
All other parameter are the same as in Figs.\ 2--6.}
\label{fig:realdata}
\end{figure}

Detailed information on our experiment can
be found elsewhere \cite{Youn06,Choi06}. In short, a few rubidium atoms
were trapped in a microscopic MOT with a diameter of a few microns
and fluorescence induced by a trap laser was measured in photon
counting mode. A raw experimental data, a segment of which is
shown in Fig.\ \ref{fig:realdata}(a), was taken with a bin time of
0.20 s. The atom-number correlation time $\tau$ was measured to
be 23 s, resulting in $\Delta t/\tau$ of  0.0086. 
Distributions with larger values of $\Delta t/\tau$ are derived
from the raw data by combining counts in neighboring time bins.

We pay close attention to Fig.\ \ref{fig:realdata}(b), where
the fit is given by Eq.\ (\ref{eq:smallt}) with each $P(S_i|m)$ given by a Gaussian distribution with a mean of $\bar{S}_m=m{\bar A}+{\bar B}$ and a variance of $\sigma_m^2$ to be evaluated below.

The average background photon number, $\bar{B}=\bar{b}\Delta t$,
and the average level spacing, $\bar{A}=\bar{a}\Delta t$, are 627
and 653, respectively, obtained from the experiment. By counting
the individual loading and loss events in the time trace of
fluorescence as shown in Fig.\ \ref{fig:realdata}(a), one can
measure the loading rate $R$ and the one- and two-atom loss rates
$\Gamma_1$ and $\Gamma_2$, respectively, and the results are
$R$=0.080 s$^{-1}$, $\Gamma_{1}$=0.043 s$^{-1}$,
$\Gamma_{2}$=0.0056 s$^{-1}$. The detailed information on
experiments to measure these rates can be
found elsewhere \cite{Choi06}. From these parameters we obtain $\bar{N}$=1.6. The Poisson distribution for this actual $\bar N$ is used for $P_{N=m}$ in Eq.\ (\ref{eq:smallt}) for the fit. Note that the only fitting parameter is then the variance $\sigma_m^2$, which can be decomposed into
\begin{equation}
\sigma_m^2=\sigma_B^2+\sigma_S^2
\end{equation}
where $\sigma_B^2$ and $\sigma_S^2$ are the variances of background and signal counts, respectively.

The background counts are mostly due to scattered light of trap
and repump lasers of MOT. Due to long-term power fluctuations, the
mean value of background counts also fluctuates, and as a result,
the width of the zero-atom peak in the signal distribution
$P(S_i)$ becomes larger than that of a Poissonian distribution. In fact, the background variance $\sigma_B^2$ was measured to be 2570$\pm$180, about 4.5 times larger than the mean count $\bar B$.

If we assume that the fluorescence counts follow
Poisson statistics, the variance $\sigma_m^2$ can be modeled as
\begin{equation}
\sigma_{m}^{2}=\sigma_{B}^{2}+m\sigma_{A}^2=(2570\pm 180)+m\bar{A}
\label{eq:variance_m_1}
\end{equation}
where $\sigma_A^2$ is the variances of one-atom fluorescence, and
it is assume that the fluorescence from one atom is statistically
independent from that of another atom. However, as shown in Fig.\
\ref{fig:realdata}(c), the observed variances of individual peaks
are not well fit by the above formula. Rather they are well fit by
an empirical formula given by
\begin{equation}
\sigma_{m}^{2}=(2570 \pm 180)+(2280\pm 190) m\;, \label{eq:variance_m_2}
\end{equation}
the slope of which is about four times larger than that of Eq.\
(\ref{eq:variance_m_1}).

The fact that the variance is still linear in $m$ indicates that the fluorescence from one atom is still statistically independent from that of another atom. This observation excludes, as a source of the increased variance, the fluorescence dispersion due to power fluctuation of trap and probe lasers, mechanical vibrations and similar technical noises since they all have to induce correlated fluctuations in the signals of individual atoms and thus proportional to $m^2$.

One possible reason for this increase variance is the motional effect of individual atoms. The atoms move independently from each other inside the MOT. Due to the spatially inhomogeneous magnetic field, atoms experience different Zeeman shifts and thus their upper level populations vary in time differently and independently from one atom to another. This variation can give rise to the observed increased variance in fluorescence counts.

Such motional effect might be observed in the second order
correlation function of the fluorescence in the long time limit.
In the short time limit, comparable to the life time of the atom
(tens of nanosecond), antibunching characteristics of the
resonance fluorescence will be dominant effect. But in the long
time limit, much longer than the atomic life time and comparable
to the characteristic time ($\sim$ millisecond) of atomic motion
in the trap, an oscillatory feature would appear in the second
order correlation function. The detailed study on this phenomena
is beyond the scope of this paper and left for the future work.

\section{Conclusion}
We have derived analytic expressions for signal distribution
$P(S_i)$ of fluorescence photo-counts from a few-atom MOT and
compared the results with Monte Carlo simulations and experimental
data. The signal distribution strongly depends on the relative
size of the bin time $\Delta t$ of photon counting with respect to
the trap decay time $\tau$. In the limit of $\Delta t \ll\tau$,
the distribution shows multiple peaks with the integrated areas of
individual peaks constituting the atom-number distribution
function $P_N$. Conversely, the stepwise fluorescence signal
corresponding to a multi-peak distribution can be regarded as a
definitive evidence of a few atoms in the trap. As $\Delta t
\ll\tau$ is increased, a broad background appears and eventually
outgrows sharp peaks corresponding to atom numbers and turns into
a single peak in the limit of $\Delta t\gg\tau$. The validity of
our derivation was confirmed by comparing the results with those
of numerical simulations including Monte Carlo simulation. These
theoretical results were then compared with experimental results.
Fluorescence photo-count distributions were observed to be
super-Poissonian, the origin of which might be due to the
statistically independent motion of atoms in the inhomogeneous
magnetic field of MOT. Our results provide necessary theoretical
background for analyzing and interpreting the fluorescence signal
of a few atom MOT and also clarify the optimum condition on the
bin time in actual experiments.

This work was supported by National Research Laboratory Grants and Korea Research Foundation Grants (KRF-2002-070-C00044, -2005-070-C00058).

\newpage

%\widetext

%{\bf List of Recommended Referees}\\
%\begin{enumerate}
%\item Wolfgang Ertmer, Institute of Quantum Optics. University of Hannover; ertmer@iqo.uni-hannover.de
%\item P. Grangier, Institute d'Optique, France; philppe.grangier@iota.u-psud.fr
%\item Deter Meschede, Institut f?r Angewandte Physik der Universit?t Bonn; meschede@iap.uni-bonn.de
%\end{enumerate}

\end{document}